\documentclass[conference, a4paper]{IEEEtran}

\IEEEoverridecommandlockouts

\usepackage[utf8]{inputenc} 
\usepackage[T1]{fontenc}
\usepackage{bm}
\usepackage{algorithmic}
\usepackage{graphicx}
\usepackage{textcomp}
\usepackage{xcolor}
\usepackage{float}
\usepackage{amsthm}
\usepackage{epstopdf}
\usepackage{bm,bbm}
\usepackage{amsfonts}
\usepackage{amssymb}
\usepackage{color}
\usepackage{multirow}
\usepackage{multicol}
\usepackage{soul,xcolor}
\usepackage{setspace}

\theoremstyle{plain}

\newcommand{\vect}[1]{\mathbf{#1}}

\def\Htran{\mbox{\tiny $\mathrm{H}$}}
\def\Ttran{\mbox{\tiny $\mathrm{T}$}}
\def\imagunit{\mathsf{j}} 

\usepackage{cite}
\usepackage[cmex10]{amsmath}

\definecolor{orange}{RGB}{0,112,192}
\usepackage{multirow}
\usepackage{array}
\usepackage[lofdepth,lotdepth]{subfig}

	\AtBeginDocument{%
		\renewcommand\tablename{TABLO}
	}
	
	\AtBeginDocument{%
		\renewcommand\abstractname{Abstract}
	}

\begin{document}

		\title{Comprehensive Analysis of Behavioral Hardware Impairments in Cell-Free Massive MIMO-OFDM Uplink: Centralized Operation }

		\author{\IEEEauthorblockN{\"Ozlem Tu\u{g}fe Demir\IEEEauthorrefmark{1}, Muhammed Selman Somuncu\IEEEauthorrefmark{2}, Ahmet M. Elbir\IEEEauthorrefmark{3}, 
				Emil Bj\"ornson\IEEEauthorrefmark{4} \thanks{\"O. T. Demir was supported by 2232-B International Fellowship for Early Stage Researchers Programme funded by the Scientific and Technological Research Council of Turkiye. E.~Bj\"ornson was supported by the Knut and Alice Wallenberg Foundation.}}    
			\IEEEauthorblockA{\IEEEauthorrefmark{1}Department of Electrical-Electronics Engineering, TOBB ETÜ, Ankara, Turkiye (ozlemtugfedemir@etu.edu.tr) \\
				\IEEEauthorrefmark{2}ASELSAN A.Ş., Ankara, Turkiye (mssomuncu@aselsan.com) \\
				\IEEEauthorrefmark{3}Department of Electrical and Electronics Engineering, Istinye University, Istanbul, Turkiye (ahmetmelbir@ieee.org)\\
				\IEEEauthorrefmark{4} Department of Computer Science, KTH Royal Institute of Technology, Stockholm, Sweden (emilbjo@kth.se)
		}}

		\maketitle

		\begin{abstract}
			Cell-free massive MIMO is a key 6G technology, offering superior spectral and energy efficiency. However, its dense deployment of low-cost access points (APs) makes hardware impairments unavoidable. While narrowband impairments are well-studied, their impact in wideband systems remains unexplored. This paper provides the first comprehensive analysis of hardware impairments—such as nonlinear distortion in low-noise amplifiers, phase noise, in-phase/quadrature imbalance, and low-resolution analog-to-digital converters—on uplink spectral efficiency in cell-free massive MIMO. Using an OFDM waveform and centralized processing, APs share channel state information for joint uplink combining. Leveraging Bussgang decomposition, we derive a distortion-aware combining vector that optimizes spectral efficiency by modeling distortion as independent colored noise.
			
		\end{abstract}
		\begin{IEEEkeywords}
			cell-free massive MIMO, hardware impairments, low-noise amplifiers, nonlinearities.
		\end{IEEEkeywords}

		\IEEEpeerreviewmaketitle
		
		\IEEEpubidadjcol

		\section{Introduction}
		Cell-free massive MIMO is a post-cellular architecture where densely deployed access points (APs) collaboratively serve multiple user equipments (UEs) over the same time-frequency resources \cite{Ngo2017b,cell-free-book,ngo2024ultradense}. By eliminating conventional cell boundaries, this distributed antenna system enables seamless multi-point transmission and significantly enhances spectral and energy efficiency, making it a key enabler for 6G.
		
		Despite the large number of serving antennas, individual APs are low-cost with only a few antennas, making them suitable for energy-efficient ultra-dense deployments. However, this cost-effectiveness introduces inevitable radio frequency (RF) hardware impairments \cite{8891922}, including nonlinear distortions in low-noise amplifiers (LNAs), low-resolution analog-to-digital converters (ADCs), in-phase/quadrature imbalance (IQI), and phase noise. Understanding their impact is crucial for evaluating practical system performance.
		
		This paper presents an analysis of cell-free massive MIMO-OFDM (orthogonal frequency division multiplexing) with low-cost hardware, focusing on deterministic behavioral models for hardware impairments. While most prior works rely on stochastic additive models for analytical tractability \cite{10225319,9103262}, these may not accurately capture real transceiver impairments in OFDM systems. Deterministic models, on the other hand, better reflect nonlinearities with fewer parameters and no dependency on a specific RF front-end \cite{Schenk2008a}.
		
		Prior studies on hardware impairments in cell-free massive MIMO largely consider narrowband models, with exceptions such as \cite{10437146}, which analyzes phase noise in OFDM, and \cite{9576714}, which addresses power amplifier nonlinearities. However, no work has comprehensively examined the combined effects of LNA distortion, phase noise, IQI, and ADC quantization in a unified manner. This paper fills this gap by providing a performance analysis for uplink centralized operation, deriving an achievable spectral efficiency (SE) expression using the Bussgang decomposition, and obtaining the optimal receive combining vector while modeling distortion noise as independent colored noise.

		\section{System Model}
		An uplink cell-free massive MIMO with frequency-selective channel is considered. There are $L$ APs and $K$ user equipments (UEs) that are randomly distributed in the coverage area. All UEs have a single antenna while each AP is equipped with $N$ antennas. The complex baseband equivalent frequency-selective channel from each UE to each AP is modeled as a finite impulse response (FIR) filter with $R$ equally sample-spaced taps.\footnote{Depending on the varying delay spreads between different APs and UEs, the number of taps will be different in practice. We can set the maximum of the delay taps as the common $R$ without loss of generality. To simplify the analysis, a common delay tap assumption is also adopted \cite{Jin2020}.}  The sampling period is $T_s=1/B$ where $B$ is the total bandwidth of $M$ OFDM subcarriers with $B/M$ subcarrier spacing. 
		
		The $r$-th tap of the channel from UE $k$ to AP $l$ is denoted by ${\bf h}_{kl}[r]\in \mathbb{C}^{N}$, for $r=0,\ldots,R-1$. The number of subcarriers satisfies $M>R$, and there are $M+R-1$ samples in each OFDM symbol with a cyclic prefix (CP) length of $R-1$. The channels are assumed to be constant and take an independent realization in each coherence block consisting of $n_{\rm{coh}}$ OFDM symbols \cite{Pitarokoilis2016a}. The coherence time  is then $T_{\rm{coh}}=n_{\rm{coh}}(M+R-1)T_s$. The channels are modeled using correlated Rayleigh fading, i.e.,  $\vect{h}_{kl}[r]\sim\mathcal{N}_{\mathbb{C}}({\bf 0},{\bf R}_{klr})$ and they are independent among different delay taps. The covariance matrix ${\bf{R}}_{klr}\in \mathbb{C}^{N\times N}$ specifies  the spatial correlation of the channel $\vect{h}_{kl}[r]$ between the antennas of AP $l$ and the large-scale effects such as pathloss and shadowing \cite{Mollen2016a,Ucuncu2020,Jin2020}. The spatial correlation matrices are fixed throughout the communication. We assume perfect channel state information (CSI) is available at the APs.

		All the APs are connected to a central processing unit (CPU) via fronthaul connections. The CPU is the center point with large computational resources and to focus on the hardware impairments at the APs that are low-complexity radio units, we assume the fronthaul connections are error- and latency-free.  
		There are mainly two types of uplink operation in cell-free networks. The first one is the centralized operation, which is the most advanced level, wherein all the processing regarding channel estimation and payload data detection are performed in the CPU \cite[Sec. 5.1]{cell-free-book}. On the other hand, in the distributed operation \cite[Sec. 5.2]{cell-free-book}, APs first estimate the channels and obtain a soft estimate of each UE's data locally. Then, these soft estimates are gathered at the CPU for final decoding of the data. In this paper, we focus on centralized operation.

		Let us focus on one of the OFDM symbols that are allocated for uplink data transmission. The frequency-domain uplink symbol of UE $k$ at the $m$th subcarrier is given by $\overline{s}_k[m]$.  The baseband equivalent signal transmitted from UE $k$ at the $q$th time-domain sample is given by the $M$-point inverse discrete Fourier transform (DFT) of the sequence $\overline{s}_k[m]$, i.e.,
		\begin{align}
			s_k[q] =
			\frac{1}{\sqrt{M}}\sum_{m=0}^{M-1}\overline{s}_k[m]e^{\imagunit \frac{2\pi q m}{M}}, 
		\end{align}
		where $q=-(R-1),\ldots,M-1$, and we assume that a CP of length $R-1$ is appended to the time-domain samples as in pilot transmission. The baseband equivalent of the distortion-free received signal at the $N$ antennas of AP $l$ is given as
		\begin{align}
			\vect{y}_{l}'[q]=\sum_{k=1}^K\sum_{r=0}^{R-1}\vect{h}_{kl}[r]\sqrt{p_k}s_k[q-r] + \vect{w}_l'[q],  \label{eq:received-data}
		\end{align}
		for $q=0,\ldots,M-1$. Each antenna RF signal is first passed through the LNA dedicated to that antenna branch and using the third-order quasi-memoryless model, the distorted signal at the output of the $n$-th LNA is given by
		\begin{align}\label{eq:y-check}
			y_{l,n}''[q]=b_{l,n,1}y_{l,n}'[q]+\frac{b_{l,n,2}}{\mathbb{E}\{|y_{l,n}'[q]|^2\}}\left\vert y_{l,n}'[q] \right \vert^2y_{l,n}'[q],
		\end{align} 
		where $y_{l,n}'[q]\in \mathbb{C}$ and $y_{l,n}''[q]\in \mathbb{C}$ denote the $n$-th element of the received signal $\vect{{y}}_l'[q]$ and the output of LNA $\vect{y}_l''[q]$, respectively. We assume long-term automatic gain control is used at the AP receiver \cite{Bjornson2019e,Demir2020}. The hardware parameters $b_{l,n,1} \in \mathbb{C}$ and $b_{l,n,2} \in \mathbb{C}$ are specific to the $n$-th LNA.  Third-order polynomial with odd-order terms are widely used to model the compression effect of the power amplifiers since the third order intercept point that is related
		to the third-order term is a common quality measure for the distortion in RF amplifiers \cite{Ronnow2019}. Quasi-memoryless modeling is meaningful when the bandwidth of the transmit signal is sufficiently low compared to the total bandwidth of the power amplifier \cite{Jacobsson2018, Schenk2008a}.

		In the next stage, the outputs of the LNAs are down-converted by mixing them with a carrier signal generated by a local oscillator (LO).\footnote{Separate LOs can be used for individual antennas or antenna subsets \cite{Bjornson2014a} alternatively. In this paper, we will assume common LO in accordance with \cite{Jacobsson2018}. However, a similar analysis to that is done in this paper can be used for other structures.} Following the model in \cite{Moghaddam2019,Fettweis2005a} that assume non-frequency selective IQI and residual phase noise, the down-converted signal in front of the ADC at the $n$-th antenna of AP $l$ can be expressed as
		\begin{align} \label{eq:y_m}
			y_{l,n}'''[q]=e^{\imagunit\psi_l[q]}\left(y_{l,n}''[q]+\alpha_{l,n}y_{l,n}''^{*}[q] \right),
		\end{align}
		where $\alpha_{l,n}\in \mathbb{C}$ is the complex weight for the IQI in the $n$th receiver branch of AP $l$. Unlike other parameters, $e^{\imagunit\psi_l[q]}$ is the residual phase noise that is common to each antenna branch due to the single shared LO. This modeling is meaningful since the phase noise results from the random phase variations from the LO imperfections \cite{Jacobsson2018}.  Usually, $\psi_l[q]$ is modeled by the samples of a first-order recursive stationary process as in \cite{Jacobsson2018} and it is independent of other signals. 
		In this case, $\psi_l[q]$ is zero-mean Gaussian random variable with variance  $\sigma^2_{\psi_l}$. We note that the phase noise process has a memory and the correlation between adjacent samples of  $\psi_l[q]$ is $0<\lambda_{\psi_{l}}<1$. In this paper, to focus on sample-based signal detection, we will not consider the dynamic structure of the phase noise, instead will focus on current sample statistics. This is motivated by the slowly time varying characteristics of the phase noise. The variance $\sigma^2_{\psi_l}$ equals $(2\pi\beta T_s)/(1-\lambda_{\psi_{l}})^2$ where $\beta>0$ is a phase-noise innovation rate parameter and $T_s$ is the sampling period.
		
		Let $Q(\cdot)$ denote the quantization function with $b$ bits and $D=2^b$ quantization levels, $l_1,\cdots,l_D$,
		and the quantization function is given by
		\begin{align} \label{eq:quantization-function}
			Q(x)=l_d \quad \text{if } x\in[ \upsilon_{d-1}, \upsilon_d ), \quad d=1,\cdots,D,
		\end{align}
		where $-\infty=\upsilon_0<\upsilon_1<\cdots<\upsilon_{D-1}<\upsilon_D=\infty$ are the thresholds. By applying the scalar quantization $Q(\cdot)$ to the real and imaginary parts of  $\vect{y}_l'''[q]$, we obtain the complex-valued quantized signal $\vect{y}_l[q] \in \mathbb{C}^{N}$ is given by
		\begin{align} \label{eq:tilder}
			\vect{y}_l[q]=Q\left(\Re(\vect{y}_l'''[q])\right)+\imagunit Q\left(\Im(\vect{y}_l'''[q])\right).
		\end{align}

		\section{Distortion-Aware Linear Receivers}

		Taking the $M$-point DFT of the sequence $\vect{y}_l[q]$, we obtain
		\begin{align} \label{eq:Bussgang-channel-frequency}
			\overline{\vect{y}}_l[m]=&\frac{1}{\sqrt{M}}\sum_{q=0}^{M-1}\vect{y}_l[q]e^{-\imagunit \frac{2\pi q m}{M}}
		\end{align}
		for $m=0,\ldots,M-1$. Now we can consider data detection separately at each subcarrier. In the following, we will go through the centralized cell-free operation.
		
		\subsection{Centralized Cell-Free Operation}
		In the centralized operation, each AP $l$ sends its distorted frequency-domain signals $\overline{\vect{y}}_l[m]$ to the CPU. The overall concatenated received signal at the $m$th subcarrier is
		\begin{equation}
			\begin{gathered}
				\overline{\vect{y}}[m]=\begin{bmatrix}\overline{\vect{y}}_1^{\Ttran}[m] \cdots \overline{\vect{y}}_L^{\Ttran}[m]  \end{bmatrix}^{\Ttran}\in \mathbb{C}^{LN}.
			\end{gathered}
		\end{equation}
		Defining $\overline{\vect{s}}[m]=[\overline{s}_1[m] \ \cdots \ \overline{s}_K[m]]^{\Ttran}$ and applying Bussgang decomposition for $\overline{\vect{y}}[m]$ by treating 
		$\overline{\vect{s}}[m]$ as the input signal \cite{Demir2021}, we obtain
		\begin{align} \label{eq:Bussgang-data}
			\overline{\vect{y}}[m]=\underbrace{\mathbb{E}\left\{ \overline{\vect{y}}[m]\overline{\vect{s}}^{\Htran}[m]\right\}}_{\triangleq\vect{B}[m]}\overline{\vect{s}}[m]+\overline{\bm{\eta}}[m], \quad q=0,\ldots,M-1,
		\end{align}
		where we used $\mathbb{E}\left\{\overline{\vect{s}}[m]\overline{\vect{s}}^{\Htran}[m]\right\}=\vect{I}_K$.
		The expectations are with respect to the random data symbols. We call $\vect{B}[m]\in \mathbb{C}^{LN\times K}$ the Bussgang gain matrix. The distortion noise $\overline{\bm{\eta}}[m]\in\mathbb{C}^{LN}$ is uncorrelated with $\overline{\vect{s}}[m]$ by construction.

		Before data detection takes place, the CPU applies the combining denoted in vector form by $\vect{v}_{k,m}$, i.e., $\widehat{\overline{\vect{s}}}_k[m]=\vect{v}_{k,m}^{\Htran}\overline{\vect{y}}[m]$, resulting
		\begin{align}
			\widehat{\overline{\vect{s}}}_k[m]&= \vect{v}_{k,m}^{\Htran}\vect{b}_k[m]\overline{s}_k[m] +
			\sum_{\substack{i=1 \\ i \neq k}}^K\vect{v}_{k,m}^{\Htran}\vect{b}_i[m]\overline{s}_i[m] \nonumber\\
			&\quad +\vect{v}_{k,m}^{\Htran}\overline{\bm{\eta}}[m],
		\end{align}
		where $\vect{b}_k[m]$ denotes the $k$-th column of the Bussgang gain matrix $\vect{B}[m]$. Since distortion plus interference is uncorrelated with the desired signal, we can use the worst-case uncorrelated
		additive noise theorem \cite{Hassibi2003a,Bjornson2019e} to obtain a lower bound to the channel capacity. The resulting SE of UE $k$ at subcarrier $m$ is given as
		\begin{align}
			&\mathrm{SE}_{k,m} =  
			\mathbb{E}\left\{\log_{2}{(1+\gamma_{k,m})} \right\}
		\end{align}
		where $\gamma_{k,m}$ corresponds to the signal to interference-plus-noise ratio (SINR) as
		\begin{align}
			\gamma_{k,m} =
			\frac
			{\left| \vect{v}_{k,m}^{\Htran}
				{\vect{b}}_k[m] \right|^2}
			{\sum_{\substack{i=1 \\ i \neq k}}^K\left| \vect{v}_{k,m}^{\Htran}
				{\vect{b}}_{i}[m] \right|^2
				+\vect{v}_{k,m}^{\Htran}\vect{C}_{\overline{\eta}\overline{\eta}}\vect{v}_{k,m}},
		\end{align}
		where $\vect{C}_{\overline{\eta}\overline{\eta}}=\mathbb{E}\{\overline{\bm{\eta}}[m]\overline{\bm{\eta}}^{\Htran}[m]\}$. This matrix is hard to compute in closed-form. Instead, it can be computed using Monte Carlo trials. 
		The optimal combining vector that maximizes  $\gamma_{k,m}$ is given as \cite{Bjornson2019e}
		\begin{align}
			\vect{v}_{k,m}^{\star}= \left( \sum_{\substack{i=1 \\ i \neq k}}^K
			{\vect{b}}_i[m]\vect{b}_i^{\Htran}[m]+ \vect{C}_{\overline{\eta}\overline{\eta}}  \right)^{-1}\vect{b}_k[m]. \label{eq:distortion-aware}
		\end{align}
		Inserting this selection of distortion-aware optimal receiver combiner $\vect{v}_{k,m}^{\star}$ into the  effective SINR  $\gamma_{k,m} $, we get 
		\begin{align}
			\mathrm{SE}_{k,m} =  
			\mathbb{E}\left\{ \vect{b}_k^{\Htran}[m]\left( \sum_{\substack{i=1 \\ i \neq k}}^K
			{\vect{b}}_i[m]\vect{b}_i^{\Htran}[m]+ \vect{C}_{\overline{\eta}\overline{\eta}}  \right)^{-1}\vect{b}_k[m]\right\}. 
		\end{align}

		\section{Numerical Experiments}
		In this section, we will quantify the centralized uplink SE performance of cell-free massive MIMO-OFDM under hardware impairments. We consider  $L=16$ randomly deployed APs in a $0.5\times 0.5$\,km$^2$ area. Each AP has $N=4$ antennas and there are $K=10$ randomly dropped UEs in this network area. The path loss in decibel scale is given according to Urban Microcell Street Canyon model as $-32.4-20\log_{10}(f_c)-31.9\log_{10}(d)+\mathcal{N}(0,8.2^2)$ where $d$ is the distance between two nodes in meters, $f_c=7.5$\,GHz is the carrier frequency, and 8.2 is the variance of the lognormal shadow fading \cite[Table 7.4.1-1]{3GPP5G}. There is a 10\,m height difference between the APs and UEs. The number of subcarriers and channel taps are $M=256$ and $R=6$, respectively. The subcarrier spacing is $15$\,kHz. The noise variance is computed for the bandwidth of $B=256\cdot 15\,\text{kHz}= 3.84$\,MHz, and a noise figure of $7$\,dB, i.e., $\sigma^2=-101.16$\,dBm. The Saleh-Valenzuela model \cite{Saleh1987} is used to model the power delay profile with $\Gamma=\gamma=2$ and 5 clusters \cite[Eqn. (7.52)]{bjornson2024introduction}. The azimuth and elevation angles of each cluster are generated uniformly randomly from the $40^{\circ}$ neighborhood of the nominal line-of-sight angles, and correlation matrices $\vect{R}_{klr}$ are generated according to the power delay profiles and uniform linear array (ULA) response vectors. The uplink transmit power is $p_k=0.1$\,W $\forall\,k$. The uniform quantizer is used with $q=2$ bits. Other hardware impairment parameters are given as follows unless otherwise stated. The third-order non-linear distortion parameters are $b_{l,n,1}=1.065$ and $b_{l,n,2}=-0.028$ \cite{Jacobsson2018}. The phase noise parameters are set as $\lambda_{\varphi_l}=0.99$, $T_s=1/B$, and $\beta=10^3$ \cite{Jacobsson2018}. Finally, $\alpha_{l,n}=0.18e^{\imagunit 0.1\pi}$ \cite{Moghaddam2019}.

		\begin{figure}[t] 
			\centering
			\includegraphics[width=0.5\textwidth, trim=0.8cm 0.2cm 1cm 0.2cm, clip]{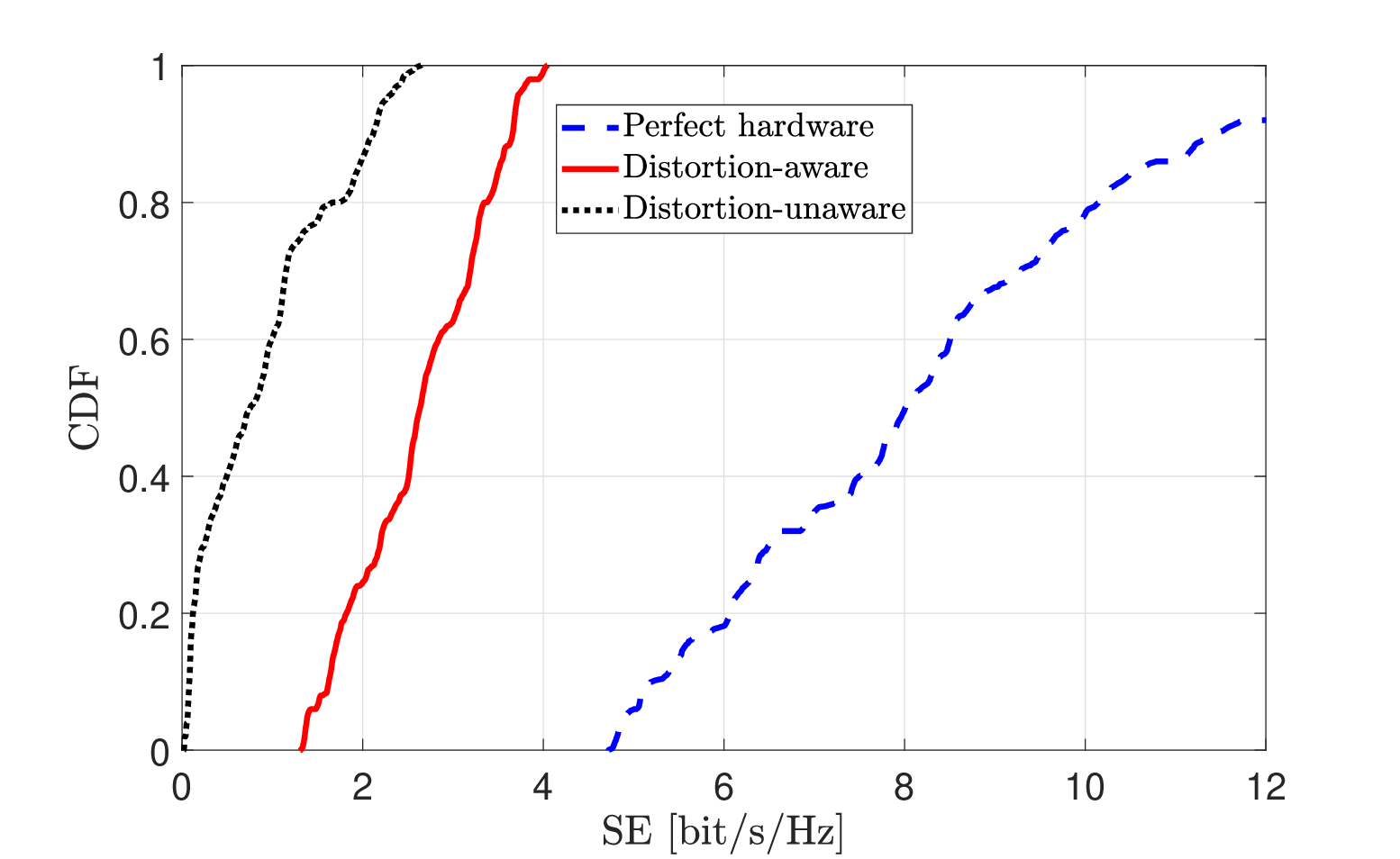} 
			\vspace{-2mm}
			\caption{The CDF of UE SEs when considering perfect hardware and imperfect hardware with different combiners. }
			\label{fig:1}
			\vspace{-2mm}
		\end{figure}
		
		\begin{figure}[t] 
			\centering
			\includegraphics[width=0.5\textwidth, trim=0.8cm 0.2cm 1cm 0.2cm, clip]{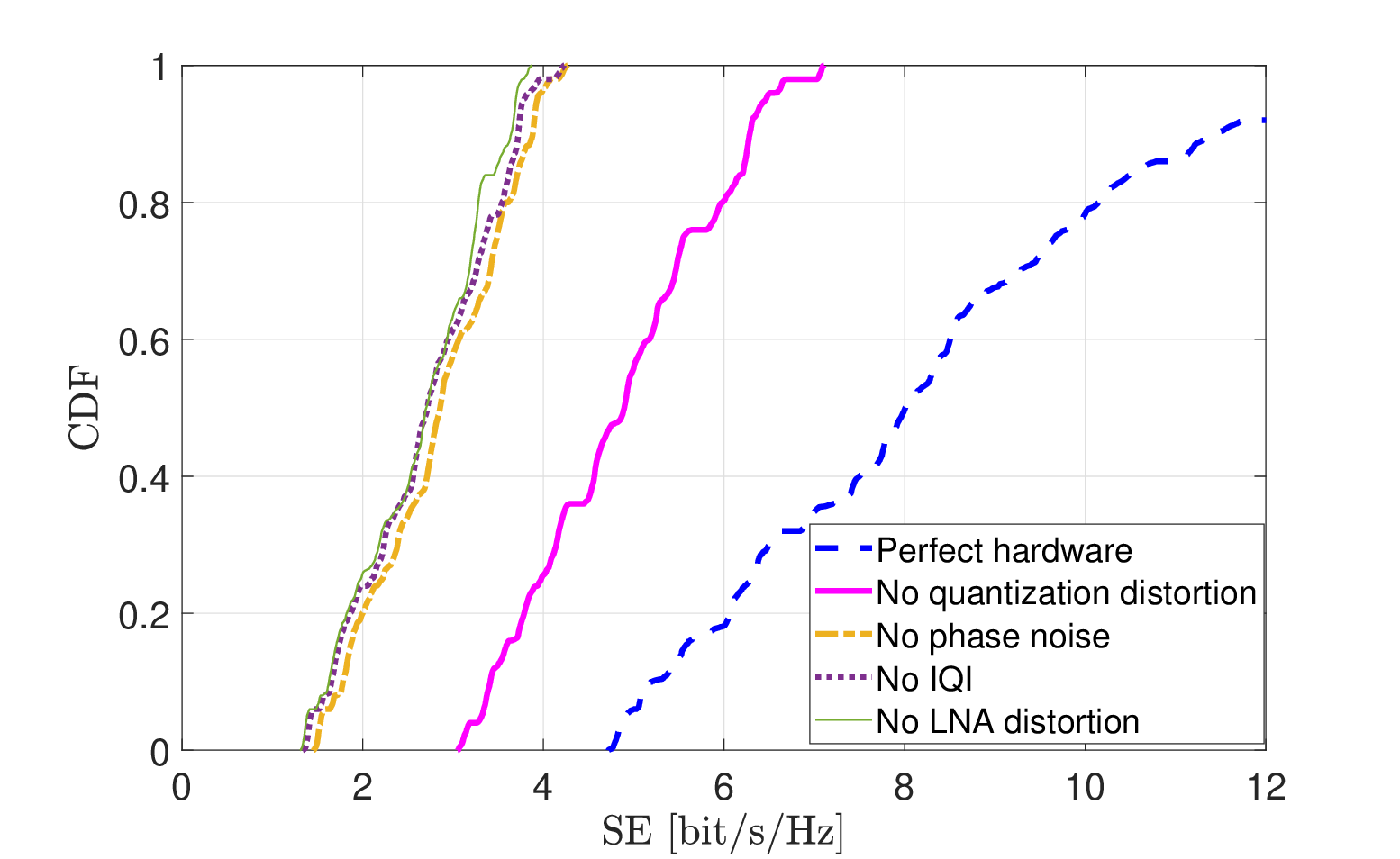} 
			\vspace{-2mm}
			\caption{The CDF of UE SEs when the effect of different hardware impairments are eliminated. }
			\label{fig:2}
			\vspace{-2mm}
		\end{figure}

		In Fig.~\ref{fig:1}, we present the cumulative distribution function (CDF) of the SE for three cases:
		i) perfect hardware—an ideal scenario without any hardware impairments, utilizing the optimal receive combining vector; ii) imperfect hardware with distortion-aware combining—where the receive combining vector is optimized according to \eqref{eq:distortion-aware} to account for hardware impairments; and
		iii) imperfect hardware with distortion-unaware combining—where the receive combining vector is designed for perfect hardware, neglecting hardware impairments. As shown in the figure, the distortion-aware combining vector significantly outperforms the distortion-unaware one. However, hardware impairments overall have a detrimental effect on SE compared to the perfect hardware case.
		
		To assess the relative impact of different hardware impairments, we selectively remove each impairment while keeping the others active in Fig.~\ref{fig:2}. The curves corresponding to the removal of phase noise, IQI, and LNA distortion exhibit similar behavior. However, a substantial performance improvement is observed when the adverse effects of limited-resolution ADC quantization are eliminated. This indicates that quantization noise has the most severe impact on SE performance compared to other hardware impairments.

		\section{Conclusions}
		This paper presents a comprehensive analysis of various behavioral hardware impairments in cell-free massive MIMO-OFDM under uplink centralized operation. We begin with the OFDM system model and incorporate the effects of LNA distortion, phase noise, IQI, and low-resolution ADCs. An achievable SE expression is derived based on the Bussgang decomposition. Furthermore, we obtain the optimal receive combining vector that maximizes the uplink SE.
		
		Simulation results reveal that hardware impairments cause significant performance degradation compared to the perfect hardware case. While the distortion-aware combining vector outperforms its distortion-unaware counterpart, a substantial performance gap remains relative to the ideal hardware scenario. Among the considered impairments, low-resolution ADCs have the most pronounced impact on SE.

		\bibliographystyle{IEEEtran}
		\bibliography{refs}

\begin{thebibliography}{10}
\providecommand{\url}[1]{#1}
\csname url@samestyle\endcsname
\providecommand{\newblock}{\relax}
\providecommand{\bibinfo}[2]{#2}
\providecommand{\BIBentrySTDinterwordspacing}{\spaceskip=0pt\relax}
\providecommand{\BIBentryALTinterwordstretchfactor}{4}
\providecommand{\BIBentryALTinterwordspacing}{\spaceskip=\fontdimen2\font plus
\BIBentryALTinterwordstretchfactor\fontdimen3\font minus
  \fontdimen4\font\relax}
\providecommand{\BIBforeignlanguage}[2]{{%
\expandafter\ifx\csname l@#1\endcsname\relax
\typeout{** WARNING: IEEEtran.bst: No hyphenation pattern has been}%
\typeout{** loaded for the language `#1'. Using the pattern for}%
\typeout{** the default language instead.}%
\else
\language=\csname l@#1\endcsname
\fi
#2}}
\providecommand{\BIBdecl}{\relax}
\BIBdecl

\bibitem{Ngo2017b}
H.~Q. Ngo, A.~Ashikhmin, H.~Yang, E.~G. Larsson, and T.~L. Marzetta,
  ``Cell-free {Massive} {MIMO} versus small cells,'' vol.~16, no.~3, pp.
  1834--1850, 2017.

\bibitem{cell-free-book}
\BIBentryALTinterwordspacing
{\"{O}}.~T. Demir, E.~Bj\"{o}rnson, and L.~Sanguinetti, ``Foundations of
  user-centric cell-free massive {MIMO},'' \emph{Foundations and Trends® in
  Signal Processing}, vol.~14, no. 3-4, pp. 162--472, 2021. [Online].
  Available: \url{http://dx.doi.org/10.1561/2000000109}
\BIBentrySTDinterwordspacing

\bibitem{ngo2024ultradense}
H.~Q. Ngo, G.~Interdonato, E.~G. Larsson, G.~Caire, and J.~G. Andrews,
  ``Ultradense cell-free massive {MIMO} for {6G}: Technical overview and open
  questions,'' \emph{Proceedings of the IEEE}, 2024.

\bibitem{8891922}
H.~Masoumi and M.~J. Emadi, ``Performance analysis of cell-free massive {MIMO}
  system with limited fronthaul capacity and hardware impairments,'' \emph{IEEE
  Transactions on Wireless Communications}, vol.~19, no.~2, pp. 1038--1053,
  2020.

\bibitem{10225319}
Y.~Zhang, W.~Xia, H.~Zhao, G.~Zheng, S.~Lambotharan, and L.~Yang, ``Performance
  analysis of {RIS}-assisted cell-free massive {MIMO} systems with transceiver
  hardware impairments,'' \emph{IEEE Transactions on Communications}, vol.~71,
  no.~12, pp. 7258--7272, 2023.

\bibitem{9103262}
Y.~Zhang, M.~Zhou, Y.~Cheng, L.~Yang, and H.~Zhu, ``{RF} impairments and
  low-resolution {ADCs} for nonideal uplink cell-free massive {MIMO} systems,''
  \emph{IEEE Systems Journal}, vol.~15, no.~2, pp. 2519--2530, 2021.

\bibitem{Schenk2008a}
T.~Schenk, \emph{{RF} imperfections in high-rate wireless systems: Impact and
  digital compensation}.\hskip 1em plus 0.5em minus 0.4em\relax Springer, 2008.

\bibitem{10437146}
Y.~Wu, L.~Sanguinetti, U.~Gustavsson, A.~G. Amat, and H.~Wymeersch, ``Impact of
  phase noise on uplink cell-free massive {MIMO OFDM},'' in \emph{GLOBECOM 2023
  - 2023 IEEE Global Communications Conference}, 2023, pp. 5829--5834.

\bibitem{9576714}
Z.~Mokhtari and R.~Dinis, ``Sum-rate of cell free massive {MIMO} systems with
  power amplifier non-linearity,'' \emph{IEEE Access}, vol.~9, pp.
  141\,927--141\,937, 2021.

\bibitem{Jin2020}
S.~N. {Jin}, D.~W. {Yue}, and H.~H. {Nguyen}, ``Spectral efficiency of a
  frequency-selective cell-free massive {MIMO} system with phase noise,''
  \emph{IEEE Wireless Communications Letters}, pp. 1--1, 2020.

\bibitem{Pitarokoilis2016a}
A.~Pitarokoilis, E.~Bj\"ornson, and E.~G. Larsson, ``Performance of the massive
  {MIMO} uplink with {OFDM} and phase noise,'' vol.~20, no.~8, pp. 1595--1598,
  2016.

\bibitem{Mollen2016a}
C.~Moll\'{e}n, J.~Choi, E.~G. Larsson, and R.~W. Heath, ``Uplink performance of
  wideband {Massive} {MIMO} with one-bit {ADCs},'' vol.~16, no.~1, pp. 87--100,
  2017.

\bibitem{Ucuncu2020}
A.~B. {Üçüncü}, E.~{Björnson}, H.~{Johansson}, A.~O. {Yılmaz}, and E.~G.
  {Larsson}, ``Performance analysis of quantized uplink massive mimo-ofdm with
  oversampling under adjacent channel interference,'' \emph{IEEE Transactions
  on Communications}, vol.~68, no.~2, pp. 871--886, 2020.

\bibitem{Bjornson2019e}
E.~{Björnson}, L.~{Sanguinetti}, and J.~{Hoydis}, ``Hardware distortion
  correlation has negligible impact on ul massive {MIMO} spectral efficiency,''
  \emph{IEEE Transactions on Communications}, vol.~67, no.~2, pp. 1085--1098,
  2019.

\bibitem{Demir2020}
{\ifmmode\ddot{O}\else\"{O}\fi}.~T. Demir and
  E.~Bj{\ifmmode\ddot{o}\else\"{o}\fi}rnson, ``{Joint Power Control and LSFD
  for Wireless-Powered Cell-Free Massive MIMO},'' \emph{IEEE Trans. Wireless
  Commun.}, vol.~20, no.~3, pp. 1756--1769, Nov. 2020.

\bibitem{Ronnow2019}
D.~{Rönnow} and P.~{Händel}, ``Nonlinear distortion noise and linear
  attenuation in {MIMO} systems—theory and application to multiband
  transmitters,'' \emph{IEEE Transactions on Signal Processing}, vol.~67,
  no.~20, pp. 5203--5212, 2019.

\bibitem{Jacobsson2018}
S.~{Jacobsson}, U.~{Gustavsson}, G.~{Durisi}, and C.~{Studer}, ``Massive
  {MU-MIMO-OFDM} uplink with hardware impairments: Modeling and analysis,'' in
  \emph{2018 52nd Asilomar Conference on Signals, Systems, and Computers},
  2018, pp. 1829--1835.

\bibitem{Bjornson2014a}
E.~Bj{\"{o}}rnson, J.~Hoydis, M.~Kountouris, and M.~Debbah, ``Massive {MIMO}
  systems with non-ideal hardware: Energy efficiency, estimation, and capacity
  limits,'' vol.~60, no.~11, pp. 7112--7139, 2014.

\bibitem{Moghaddam2019}
M.~H. {Moghaddam}, S.~R. {Aghdam}, and T.~{Eriksson}, ``An additive noise
  modeling technique for accurate statistical study of residual rf hardware
  impairments,'' in \emph{2019 IEEE International Conference on Communications
  Workshops (ICC Workshops)}, 2019, pp. 1--5.

\bibitem{Fettweis2005a}
G.~Fettweis, M.~L\"{o}hning, D.~Petrovic, M.~Windisch, P.~Zillmann, and
  W.~Rave, ``Dirty {RF}: A new paradigm,'' in \emph{IEEE PIMRC}, 2005, pp.
  2347--2355.

\bibitem{Demir2021}
O.~T. {Demir} and E.~{Bjornson}, ``The {Bussgang} decomposition of nonlinear
  systems: Basic theory and {MIMO} extensions [lecture notes],'' \emph{IEEE
  Signal Processing Magazine}, vol.~38, no.~1, pp. 131--136, 2021.

\bibitem{Hassibi2003a}
B.~Hassibi and B.~M. Hochwald, ``How much training is needed in
  multiple-antenna wireless links?'' vol.~49, no.~4, pp. 951--963, 2003.

\bibitem{3GPP5G}
\emph{{5G}; Study on channel model for frequencies from 0.5 to 100 GHz (Release
  14)}.\hskip 1em plus 0.5em minus 0.4em\relax {3GPP} {TR} 38.901, Jan. 2018.

\bibitem{Saleh1987}
A.~A. Saleh and R.~A. Valenzuela, ``A statistical model for indoor multipath
  propagation,'' vol.~5, no.~2, pp. 128--137, 1987.

\bibitem{bjornson2024introduction}
E.~Bj{\ifmmode\ddot{o}\else\"{o}\fi}rnson and
  {\ifmmode\ddot{O}\else\"{O}\fi}.~T. Demir, \emph{{Introduction to Multiple
  Antenna Communications and Reconfigurable Surfaces}}.\hskip 1em plus 0.5em
  minus 0.4em\relax Now Publishers, Inc., Jan. 2024.

\end{thebibliography}

	\end{document}